\documentclass[runningheads]{llncs}
\usepackage{graphicx}
\usepackage{hyperref}
\usepackage{cite}
\newcommand{\etal}{\textit{et al.}}
\newcommand{\circled}[1]{\raisebox{.5pt}{\textcircled{\raisebox{-.9pt} {#1}}}}

\begin{document}
\title{Utilization of machine learning for the detection of self-admitted vulnerabilities}
%
%
\author{Moritz Mock
\orcidID{0009-0009-3156-6211} 
}
\authorrunning{M. Mock}
\titlerunning{Utilization of ML for the detection of self-admitted vulnerabilities}
\institute{
Faculty of Engineering\\ Free University of Bozen-Bolzano\\ Bolzano, Italy\\
\email{moritz.mock@student.unibz.it}}
\maketitle              
\begin{abstract}
\textbf{Motivation:} 
Technical debt is a metaphor that describes not-quite-right code introduced for short-term needs. Developers are aware of it and admit it in source code comments, which is called Self-Admitted Technical Debt (SATD). Therefore, SATD indicates weak code that developers are aware of.
\textbf{Problem statement:} 
Inspecting source code is time-consuming; automatically inspecting source code for its vulnerabilities is a crucial aspect of developing software. It helps practitioners reduce the time-consuming process and focus on vulnerable aspects of the source code.
\textbf{Proposal:} 
Accurately identify and better understand the semantics of self-admitted technical debt (SATD) by leveraging NLP and NL-PL approaches to detect vulnerabilities and the related SATD. Finally, a CI/CD pipeline will be proposed to make the vulnerability discovery process easily accessible to practitioners.
\keywords{Software Vulnerabilities \and Code Comments \and Self-Admitted Technical Debt \and Machine Learning \and NL-PL}
\end{abstract}
\section{Introduction}
Technical debt (TD) is a metaphor introduced by Cunningham \cite{Cunningham1992}, which describes the short-term benefits of not-quite-right code, such as being able to deliver the code faster. In the paper of Bavota \etal~\cite{Bavota16}, it was illustrated that the negative impact of TD increases over time and that the payback of technical debt should not be postponed for too long.
In 2014, the concept of self-admitted technical debt (SATD) was introduced by Potdar and Shihab \cite{Potdar2014}, which refers to comments left behind by developers to indicate that the code is not quite right. Furthermore, they have presented 62 patterns which can identify SATD.
Due to the patterns' high precision and low accuracy, in recent years, different machine learning (ML) approaches (e.g., \cite{Huang2018, Ren2019, Zhu2023}) have been employed to detect SATD. With the ML approach of \cite{Ren2019}, they were able to identify more than 700 patterns, which, according to their approach, indicate SATD.

For the part of vulnerability detection, many static analysis tools (e.g., Cppcheck
and Flawfinder)
and ML approaches (e.g., \cite{Li2021, Fu2022, Hin2022}) already exist for all kinds of programming languages (PL) and vulnerability types. To the best of our knowledge, no tool leverages both concepts by connecting them besides our own tool WeakSATD Russo \etal~\cite{Russo22}. Furthermore, the Common Weakness Enumeration (CWE\footnote{\href{https://cve.mitre.org/}{https://cve.mitre.org/}\label{note:cwe}}) 
and the Common Vulnerability and Exposure (CVE\footnote{\href{https://cwe.mitre.org/}{https://cwe.mitre.org/}\label{note:cve}})
repositories provide a large set of abstract and real-world vulnerability instances, respectively, both of which can be employed for machine-learning-related tasks.

This paper proposes to foster the understanding of SATD and its semantics in related comments. Furthermore, the investigation of the correlation between SATD and vulnerabilities. An annotated dataset for SATD and vulnerabilities will also be presented, focusing on CWE-related vulnerabilities. For which different Natural Language Processing (NLP) and Natural Language Programming Language (NL-PL) techniques are employed. The machine learner should be able to detect the vulnerability, indicate the affected line, and provide practitioners with easy-to-understand assistance on how to fix the vulnerability.
Lastly, an integration in a continuous integration pipeline (CI/CD) is foreseen.

The rest of this doctorate proposal includes state-of-the-art SATD and vulnerability detection, which is presented in Section \ref{sec:sota}. It is followed by the Section \ref{sec:proposal}, illustrating the proposal. The conclusion and future work are presented in Section \ref{sec:conclusion}, which additionally contains the timeline for the PhD.
\section{State of the art}
\label{sec:sota}
We have reviewed the existing literature according to two dimensions: (i) SATD and (ii) vulnerability detection. In the following, we give an overview of them.\\
\textbf{SATD:} For SATD comment detection, the two primary approaches advocated in the literature are pattern-based and machine learning.
Potdar and Shihab \cite{Potdar2014} presented the fundamental paper for the pattern-based approach. They defined 62 unique SATD patterns by examining 101,762 comments over four open-source projects (Eclipse, Chromium OS, ArgoUML and Apache httpd). 
Different approaches for the detection of SATD have been studied (e.g., \cite{Aiken2023, Guo2021, Huang2018, Zhu2023}). Hung \etal~\cite{Huang2018} leveraged a Multinomial Naive Bayes approach for the detection of SATD, for which they used the manually annotated dataset of 62566 Java comments, containing 4071 SATD comments, presented by Maldonado \etal~\cite{Maldonado2017}.
Ren \etal~\cite{Ren2019} proposed a Convolutional Neural Network (CNN) to detect SATD with an average F1 of 0.766 over eight open-source projects annotated by Maldonado \etal~\cite{Maldonado2017}. Furthermore, they have investigated which keywords, besides the 62 patterns from Potdar and Shihab \cite{Potdar2014}, have statistical significance for SATD identification. With that approach, they were able to propose 700 patterns for SATD detection. In the study of Guo \etal~\cite{Guo2021}, the CNN approach of Ren \etal~\cite{Ren2019} was investigated. They could not replicate the initially obtained F1 score; their result was 0.654 for the F1, effectively being 0.112 points lower than the initial findings. In the recently published study of Aiken \etal~\cite{Aiken2023}, BERT \cite{Devlin2018} was employed to predict SATD in the dataset of Maldonado \etal~\cite{Maldonado2017}. They have achieved an average F1 of 0.858 in a cross-project evaluation. Furthermore, new sets of patterns have been introduced and used for the detection of SATD, which differ from the original proposed 62 patterns (\cite{Guo2021, Ebrahimi2023, Ren2019}). Guo \etal~\cite{Guo2021} proposed a set of patterns consisting of ``TODO'', ``HACK'', ``XXX'', and ``FIXME'', which they called Matches task Annotation Tags (MAT). Their idea is that developers are particularly prone to use those four patterns to remind themselves, as those patterns are automatically highlighted in the most common Integrated Development Environments (IDSs). 
\\
\textbf{Vulnerabilitiy:} The repository of CWE provides an in-depth view of more than 400 instances of source-code-related vulnerabilities, which are held quite general and abstract without leveraging on real-world examples. Most of them contain an abstract source-code example with steps on how to mitigate the vulnerability besides the textual description. In contrast, the CVE repository is a collection of vulnerabilities taken from real-world applications. Where possible, they contain the link to the corresponding commit-ID, which fixed the vulnerability. The dataset Big-Vul presented by Fan \etal~\cite{Fan2020} provided 3754 vulnerabilities of 91 different vulnerability types. Big-Vul was used in the study of Fu \etal~\cite{Fu2022} in which LineVul was presented, a transformer-based application leveraging on CodeBERT \cite{Feng2020}, to perform line-level predictions of vulnerabilities. With their approach, they obtained an F1 score of 0.91. 
Devign was proposed by Zhuo \etal~\cite{Zhou2019}, a Graph Neural Network (GNN) for the vulnerability of the programming language C for the detection they are leveraging on the Abstract Syntax Tree representation of source code. For their approach, they manually annotated vulnerabilities from four large open-source projects: FFMpeg, Wireshark, Linux Kernel, and QEMU. The current SOTA leverages many individually created datasets, which has only changed recently with the introduction of easy-to-use datasets such as Big-Vul \cite{Fan2020} and Devign \cite{Zhou2019}. Especially the part of explanatory vulnerability detection was not explored at all. Fu \etal~\cite{Fu2022} went in this direction by implementing a tool, LineVul, which is capable of line-level vulnerability prediction. However, they lack in the part of providing an explanation of \textit{how to fix} the detected vulnerability. Chakraborty \etal~\cite{Chakraborty2022} studied the capability of state-of-the-art machine learning vulnerability detection approaches towards real-world datasets. To perform their evaluation, they used the published dataset of Devign \cite{Zhou2019} and a curated dataset from Chromium and Debian. 
They have found that the existing datasets are rather too simple and not realistic enough, so the machine learners trained on them may not achieve the expected results.
\section{Proposal}
\label{sec:proposal}
This proposal for automated self-admitted vulnerabilities detection is illustrated in the following five major steps: \circled{1} \textbf{Dataset annotation}, \circled{2} \textbf{SATD detection}, \circled{3} \textbf{vulnerability detection}, \circled{4} \textbf{combining the obtained results of SATD and vulnerability detection}, and \circled{5} \textbf{CI/CD pipeline}.\\
\circled{1} For the annotation of a dataset, our previously presented tool, WeakSATD \cite{Russo22}, will be employed, as it gives us line-level vulnerability classification, which can be leveraged in the following steps of the proposal.\\
\circled{2} The detection of self-admitted technical debt has already been explored in literature and has been recognised. However, there is some disagreement on which patterns indicate SATD. This step is split into two parts; the first step provides a better understanding of what characterises the semantics of a SATD as such and whether there is a classification of a SATD comment as an indicator of a security breach. Traditional NLP and NL-PL approaches will be used, of which the second is underrepresented in the current literature. In the second step, a machine learner will label the source code as vulnerable and generate comments, potentially SATD, for it to further emphasise its vulnerability for developers.\\
\circled{3} For the detection of vulnerabilities, transformer-based machine-learning approaches with pre-trained models for source code understanding will be employed. Additionally, the capability of such machine learners is studied for the recommendation of mitigation steps for vulnerabilities to practitioners.\\
\circled{4} Combining both SATD and vulnerability detection approaches has not yet been explored in the literature. 
We hope to increase the measured results by leveraging both concepts, SATD and vulnerability detection and to increase the efficiency of a machine learner by reducing the needed computational power.\\
\circled{5} The existing literature highly focuses on finding the best approach to detecting SATD or vulnerabilities, and only a limited number of tools are available for integrating them into the development process of source code. Various approaches can be taken for this step, such as a command line interface (CLI) or an IDE plugin. With the advent of CI/CD pipelines since the introduction of Github Actions, these have become more common due to their asynchronous nature. 
Giving practitioners the possibility to evaluate the source code detached from their development process will increase the productivity of a practitioner and reduce the vulnerabilities in the code.
\section{Conclusion and Future work}
\label{sec:conclusion}
\begin{figure}[h!t]
    \centering
    \includegraphics[width=\linewidth]{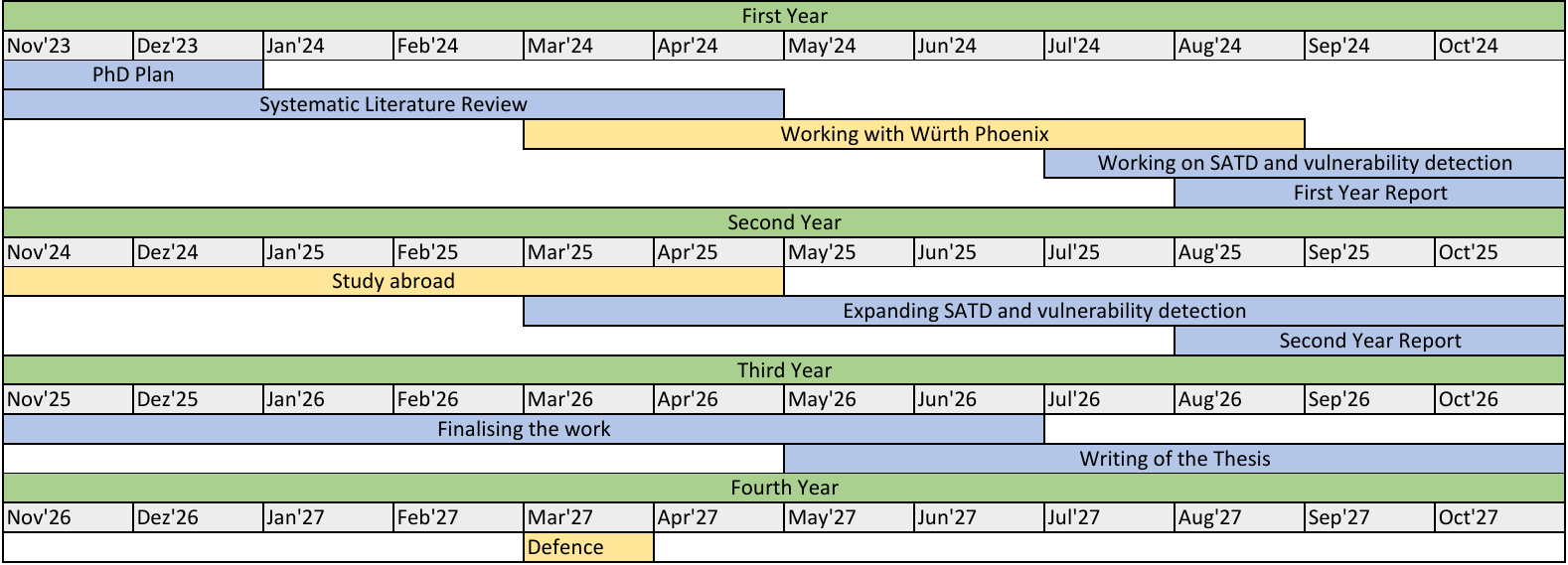}
    \caption{Timeline of the PhD study}
    \label{fig:timeline}
\end{figure}

\textbf{Conclusion:}
This doctoral proposal suggests the implementation of a CI/CD pipeline to make it as easy as possible for practitioners to access a SATD and vulnerability detection tool. Furthermore, a gap in the detection of SATD was identified with its proposed exploration of NLP and NL-PL approaches to extract a more in-depth understanding of the semantics of SATD and its connection to security breaches. Lastly, the absence of a dataset annotated at the line level for vulnerabilities was identified. 
A connection with CWE helps the practitioner find clear guidance in the explanation of the vulnerability, how to fix it, and how to mitigate it in the future can be retrieved. 
\textbf{Future work:} After each step, but especially after step \circled{2} and \circled{3} of the proposal, practitioners will be interviewed to understand how they perceive the labelling of the machine learner. As well as to showcase step \circled{5} to understand if the foreseen solution with the CI/CD pipeline is applicable in their development process. 
\textbf{Timeline:} In November 2023, the three-year PhD of Advanced-Systems Engineering started. Figure \ref{fig:timeline} illustrates the study's full timeline, with the defence in 2027. This proposal was presented at the phase of the literature review of the PhD program.

\section*{Acknowledgement}
I sincerely thank my advisor, Prof. Barbara Russo, Full Professor at the Free University of Bozan-Bolzano, for her support during my academic journey.\\

\bibliographystyle{splncs04}
\bibliography{ref.bib}

\end{document}